\documentclass[a4paper,preprint,showpacs,amssymb,pre,superscriptaddress]{revtex4}
\usepackage{amssymb}
\usepackage{graphicx}

\begin{document}

\title{Stochastic Resonance: influence of a $f^{-\kappa}$ noise
spectrum}

\author{Miguel A. Fuentes}\email{fuentesm@cab.cnea.gov.ar}
\affiliation{Centro At\'omico Bariloche and Instituto Balseiro,
8400 Bariloche, R\'{\i}o Negro, Argentina} \affiliation{Santa Fe
Institute, 1399 Hyde Park Road, Santa Fe, New Mexico 87501, USA}
\author{Horacio S. Wio}\email{wio@ifca.unican.es}
\affiliation{Centro At\'omico Bariloche and Instituto Balseiro, 8400
Bariloche, R\'{\i}o Negro, Argentina} \affiliation{Instituto
de Fisica de Cantabria, Universidad de Cantabria and CSIC,\\
Avda. Los Castros s/n, 39005 Santander, Spain}

\begin{abstract}
Here, in order to study \textit{stochastic resonance} (SR) in a
double-well potential when the noise source has a spectral density
of the form $f^{-\kappa}$ with varying $\kappa$, we have extended a
procedure, introduced by Kaulakys et al (Phys. Rev. E \textbf{70},
020101 (2004)). In order to have an analytical understanding of the
results, we have obtained an effective Markovian approximation, that
allows us to make a systematic study of the effect of such kind of
noises on the SR phenomenon. The comparison of numerical and
analytical results shows an excellent qualitative agreement
indicating that the effective Markovian approximation is able to
correctly describe the general trends.
\end{abstract}

\maketitle

\section{Introduction}
\label{intro}

\textit{Stochastic resonance} (SR) is one of the most interesting
\textit{noise-induced phenomena}, that arises from the interplay
between \textit{deterministic} and \textit{random} dynamics in a
\textit{nonlinear} system \cite{gammaitoni}. This phenomenon has
been largely studied during more than two decades due to its great
interest not only from a basic point of view but also for its
technological interest and biological implications
\cite{gammaitoni,wellens}.

Most of those studies have used white or colored noises, with a few
exceptions where more wide classes of noises were considered. For
instance, in \cite{noG} SR in systems subject to a colored and non
Gaussian noise was studied. However, there are other works more
tightly related with the present one, as indicated by the following
examples. In \cite{makra1}, the authors studied, through numerical
simulations, the behavior of the signal-to-noise ratio (SNR) gain in
a level crossing detector and a Schmidt trigger, when subject to a
colored noise composed of a periodic train pulse plus a Gaussian
$f^{-\kappa}$ noise with variable $\kappa$. Their results indicate
that the maximum of the SNR is larger for white noise, and moves
towards large noise intensities for increasing $\kappa$. In
\cite{nozaki4} experimental evidence was found that noise can
enhance the homeostatic function in the human blood pressure
regulatory system. Related with the last work, other experimental
evidence was found in \cite{soma1} that an externally applied
$f^{-1}$ noise, added to the usual white noise, contributes to
sensitizing the baroflex function in the human brain. In
\cite{kish1}, and in a model of traffic junction of a main and a
side road, it was found that the effect of a Gaussian $f^{-\kappa}$
noise with $\kappa \geq 0$ shows an overall traffic efficiency
enhancement. An enhancement of the SR phenomenon in a
FitzHugh-Nagumo model submitted to a colored noise with
$f^{-\kappa}$ for $0 \geq \kappa \geq 2$ was found in
\cite{nozaki1}. In  \cite{nozaki2} it was experimentally
demonstrated that an SR-like effect can be obtained in rat sensory
neurones with white, $f^{-1}$ and $f^{-2}$ noises, and that, under
some particular conditions, $f^{-1}$ noise can be better than white
noise to enhance neuron's response. Related to it, in \cite{nozaki3}
it was shown that it is possible to enhance the SR effect in a
FitzHugh-Nagumo model submitted to a colored noise with
$f^{-\kappa}$, and that the optimal noise variance of SR could be
minimized with $\kappa \approx 1$.

Motivated by the work of Kaulakys and collaborators
\cite{kaulakys00}, who have introduced a method to generate $f^{-1}$
noises over a wide range of frequencies (see also
\cite{kaulakys01,kaulakys02}), here we discuss how to extend such a
procedure for positive and negative values of the stochastic
variable. We also exploit this procedure to analyze the effect of a
noise spectrum of the form $f^{-\kappa}$ with varying $\kappa$, on
the SR phenomenon in a simple double-well potential. In the
following Section we present the model system to be studied and the
procedure to generate the $f^{-\kappa}$ noise. Afterwards we discuss
an effective Markovian approximation, and exploit it to study SR.
Finally we discuss the results and draw some general conclusions.

\section{The system}
\label{syst}

\subsection{Stochastic differential equations}

The starting point of our analysis is the following system of
stochastic differential equations
\begin{eqnarray}
\dot{x} & = & f(x) + g(x) y(t)    \label{x} \\
\dot{y} & = & \frac{u(y)}{\tau} +\frac{D}{\tau} v(y) \xi(t),
\label{y}
\end{eqnarray}
where $x$ is the coordinate of a particle diffusing in a double
well potential $U_o(x)=-\int^x {f(\zeta) d \zeta}=\frac{x^4}{4} -
\frac{x^2}{2},$ subject to a noise $y(t)$. The second equation
corresponds to the Langevin equation driving the noise $y(t)$,
inspired in the work of Kaulakis et al. \cite{kaulakys00}. In
this last equation we consider a new potential $V(y)=-\int^y
{u(s) ds}$, and a (white) noise $\xi(t)$ that enters in a
multiplicative form with a function $v(y)$. The last function
will be characterized by an exponent $\mu$.

We consider the following form for the function $u(y)$
\begin{eqnarray}
u(y) & = & \alpha y^3 - \beta y^5 + s(y) y^4    \label{u}
\end{eqnarray}
where $s(y)$ indicates the sign of $y$ (i.e, $-1$ if $y < 0$ and
$+1$ if $y \geq 0$). For the function $v(y)$ we adopt
\begin{eqnarray}
v(y) & = & |y|^{\mu} + c,  \label{v}
\end{eqnarray}
where both, the exponent $\mu$ and the constant $c$ are positive
($>0$).

The above indicated forms change the symmetry of the potential
$V(y)$ and, in addition, when compared with the work in
\cite{kaulakys00}, increases the range of the noise variable from
$[0, +\infty)$ to $(-\infty, +\infty)$, as is shown in Fig. 1. The
parameter $c$ allows for the random variable $y$ to adopt negative
values when $c > |y|^{\mu}$.

\begin{figure}[ht]
\includegraphics[width=11cm,angle=0]{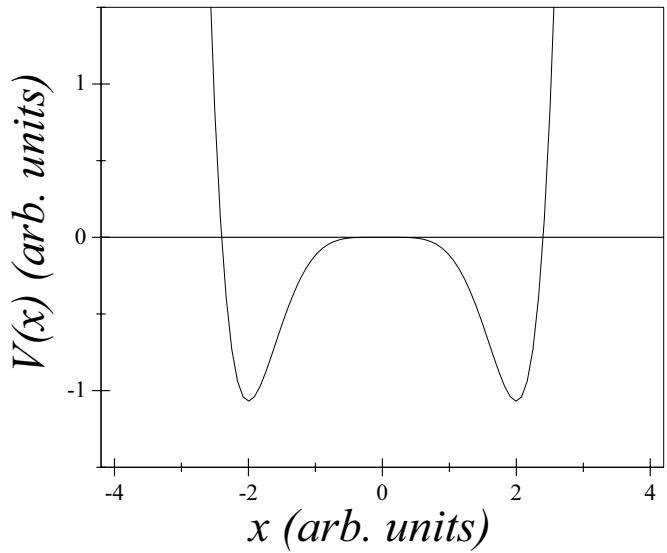}
\caption{Symmetric potential $V(y)$ as derived from Eq. (\ref{u}).
We have used the following values: $\alpha=5\times10^{-4}$ and
$\beta=\frac{1}{2}$.} \label{fig_newpot}
\end{figure}

\subsection{Characteristic of the noise variable $y$} \label{y process}

\begin{figure}[ht]
\includegraphics[width=11cm,angle=0]{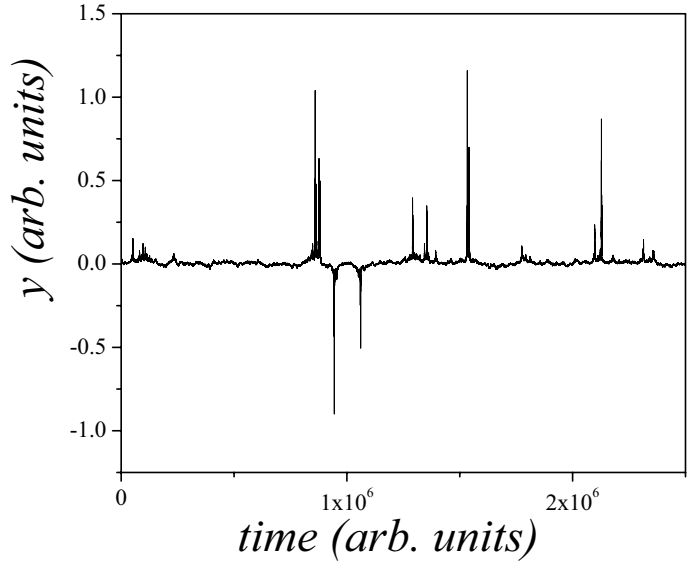}
\caption{A typical realization for Eq. (\ref{y}) using $\mu$=5/2
and $c=1 \times 10^{-4}$. The PSD of this realization shows a
$1/f$ behavior, see Fig. \ref{fig_psd25}.} \label{fig_noise25}
\end{figure}

\begin{figure}[ht]
\includegraphics[width=11cm,angle=0]{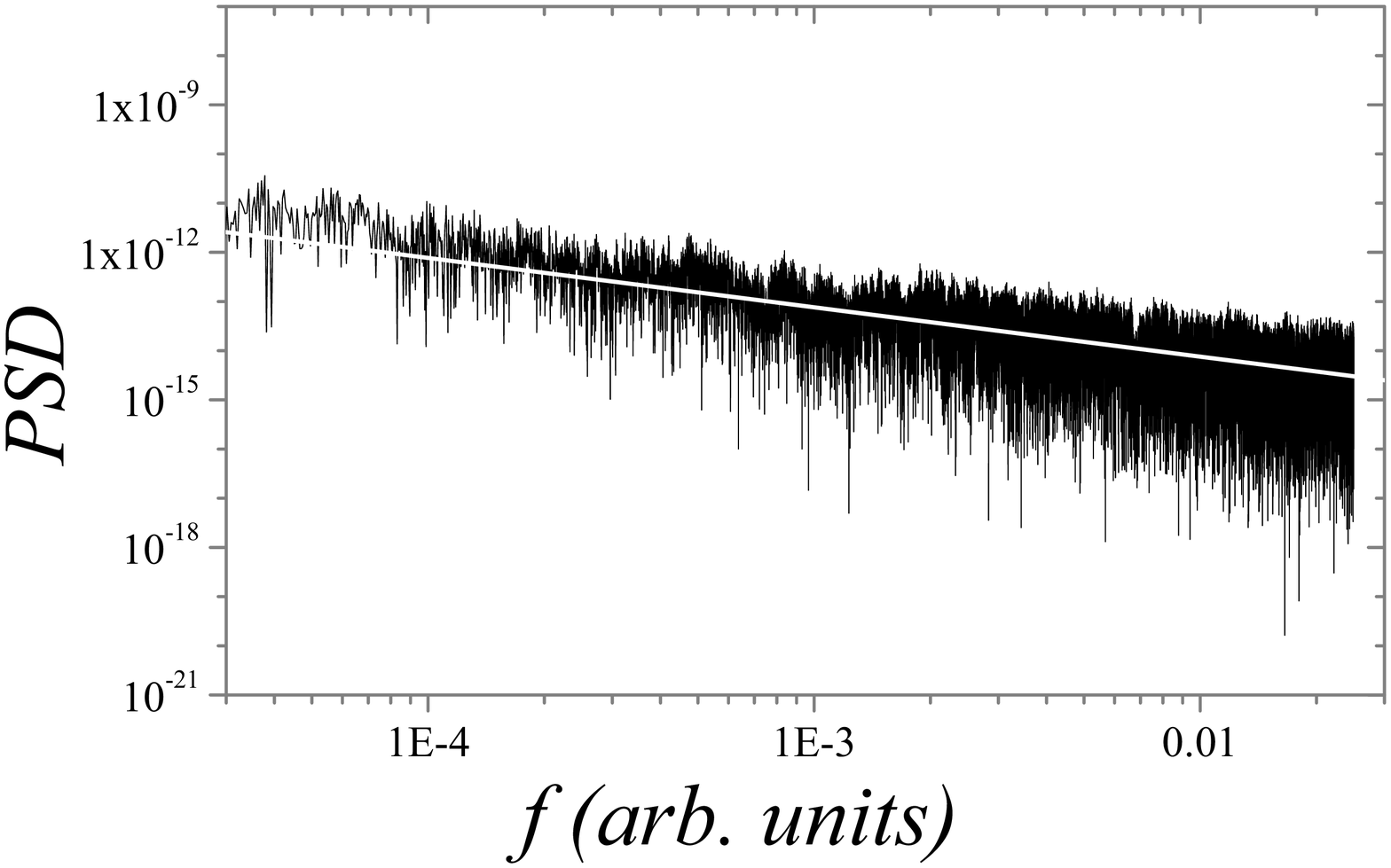}
\caption{PSD for the variable $y$, as indicated in Eq. (\ref{y}). We
used the same values of parameters as in Fig. \ref{fig_noise25}. The
white line corresponds to a linear fitting, resulting in a slope
$\kappa = - 1.004 \pm 0.005$. } \label{fig_psd25}
\end{figure}

The most relevant aspect of the process $y$ is its power spectral
density (PSD) with a $1/f$ frequency behavior. Kaulakis et al
\cite{kaulakys00} have shown that when $c=0$ and $\mu=5/2$, the
noise $y$ exhibits a $1/f$ functionality in a wide range of
frequencies. For $c>0$, but small, this property is still valid, as
we show in figures \ref{fig_noise25} and \ref{fig_psd25}.

When the exponent $\mu$ changes, the PSD behaves as $1/f^k$, with $k
< 1$ for $\mu < 5/2$. We will use this property to evaluate the
mean-first-passage-time (MFPT), and the signal-to-noise ratio (SNR).
In particular we have used $\mu = 3/2$, yielding $k \simeq 3/4$.

\section{Effective Markovian theory}
\label{ucna}

In order to be able to obtain some analytical results, we resort to
an \textit{effective Markovian approximation}, analogous to the so
called \textit{unified colored noise approximation} (UCNA)
\cite{ucna1,ucna2}. Exploiting this approach we are able to find an
effective Markovian Fokker-Planck equation (FPE) for the probability
density $P(x,t)$. The procedure is the following (the prime will
indicate derivation respect to the variable $x$).

\subsection{Adiabatic procedure}

Deriving Eq. (\ref{x}) respect to the time we have
\begin{eqnarray}
\ddot{x} & = & f'(x) \dot{x} + g'(x) \dot{x} y+ g(x) \dot{y}.
\label{x2puntos}
\end{eqnarray}
Now, assuming an adiabatic behavior, we eliminate $\ddot{x}$, and
using Eq. (\ref{y}) we obtain
\begin{eqnarray}
0 \simeq f'(x) \dot{x} + g'(x) \dot{x} \left[\frac{\dot{x} -
f(x)}{g(x)} \right] + g(x) \left[\frac{u(Z(x))}{\tau} + \frac{D
v(Z(x)) \xi(t) }{\tau}\right], \label{q2puntos1}
\end{eqnarray}
where we have defined $Z(x) = Z_0(x) + Z_1(x),$ whith $Z_0(x) = -
\frac{f(x)}{g(x)}$ and $Z_1(x) = \frac{\dot{x}}{g(x)}$. Now, we use
the following approximation
\begin{eqnarray}
u(Z(x)) \approx u(Z_0(x)) + u'(Z_0(x))\,Z_1(x) \label{aproxh}
\end{eqnarray}
and similarly for $v$
\begin{eqnarray}
v (Z) \approx v(Z_0) + v'(Z_0)\,Z_1(x).
\end{eqnarray}
Adopting now $g(x)=1$, that implies $Z_0 \equiv - f(x)$, we have
\begin{eqnarray}
0 = f'(x) \dot{x} + \frac{u(Z_0(x))}{\tau} \dot{x} + D
\frac{v(Z_0(x))}{\tau}  \xi(t) + O(\dot{x}\, \xi(t)).
\end{eqnarray}
With the above indicated results, the effective equation for the
process $x$ adopts the following form
\begin{eqnarray}
\dot{x} = - \frac{u(Z_0(x)) + D v(Z_0(x)) \xi(t)}{\tau
f'(x)+u'(Z_0(x))} = A_1(x) + B_1(x)\xi(t),
\end{eqnarray}
and, due to the polynomial character of the function $h(y)$, we can
write the following limit for $A(x)$ and $B(x)$ when $\tau \to 0$
\begin{eqnarray}
A_1(x)  &\rightarrow & \frac{Z_0(x)}{3} = A(x), \\
B_1(x)  &\rightarrow & - \frac{D v(Z_0(x))}{u'(Z_0(x))} = B(x).
\end{eqnarray}

Finally, using the above indicated approximations, the stochastic
differential equation for the process $x$ reads
\begin{eqnarray}
\dot{x} = A(x) + B(x)\, \xi(t). \label{xfinal}
\end{eqnarray}

\subsection{Fokker-Planck equation} \label{FP}

The FPE associated with the Langevin equation, Eq. (\ref{xfinal}),
is (using the Ito prescription \cite{gardi})
\begin{eqnarray}
\frac{\partial}{\partial t} P(x,t) & = & -\frac{\partial}{\partial
x} [A(x) P(x,t)]+ \frac{1}{2}\frac{\partial^2}{\partial x^2} [B^2(x)
P(x,t)]. \label{fpe}
\end{eqnarray}
As is well known, the stationary distribution of this FPE is given
by
\begin{eqnarray}
P(x) & = & \frac{N}{B(x)} \exp{\left\{ -\Phi(x) \right\} },
\end{eqnarray}
where $N$ is the normalization factor, and
\begin{eqnarray}
\Phi(x) & = & 2 \int^{x} \frac{A(\zeta)}{B(\zeta)} d\zeta.
\end{eqnarray}

\subsection{Mean-First-Passage-Time and SR} \label{FP2}

The indicated FPE and its associated stationary distribution allow
us to obtain the mean-first-passage-time (MFPT) through a
Kramers-like approximation. Using known expressions we obtain for
the MFPT \cite{gardi}
\begin{eqnarray}
T(x_0) & = & 2 \int^{x_0}_a \frac{dy}{\Psi(y)} \int_{-\infty} ^y
\frac{dz \, \Psi(z)}{B(z)}, \label{fptk}
\end{eqnarray}
where
\begin{eqnarray}
\Psi(x) & = & \exp{ \left\{ 2 \int^{x} \frac{A(\zeta)}{B(\zeta)}
d\zeta \right\} }. \label{pot_efectivo}
\end{eqnarray}

In order to study SR, as usual we introduce an external signal in
the form of a term rocking the double well potential: $U(x) = U_o(x)
+ S(t)$, with $S(t) = S_o \sin (\omega t)$ (in what follows we adopt
$\omega = 1.33 \times 10^{-5}$). Exploiting the so called
``two-state approximation" \cite{gammaitoni}, we define the SNR as
the ratio of the strength of the output signal and the broadband
noise output evaluated at the signal frequency $\omega$, obtaining
\cite{gammaitoni}
\begin{eqnarray}
SNR & \propto \left\{ \frac{1}{T} \frac{dT}{dS} \right\}_{S=0},
\label{snrtheory}
\end{eqnarray}
where the derivative of the $T$ in the above expression, as
indicated, is evaluated at $S=0$.

\section{Results and Conclusion}

We have done extensive numerical simulations of the full set of Eqs.
(\ref{x},\ref{y}) in order to obtain the SNR. The results are shown
in Fig. 4. Also, in Fig. 5 we show the SNR computed using the
effective Markovian theory, obtained through Eqs. (\ref{fptk}) and
(\ref{snrtheory}). In order to be able to compare the results we
have normalized the curves. Also, in order to have a well defined
variance of the noise process, in all the simulation we have
obtained numerically the variance of $y(t)$, as described by the Eq.
(\ref{y}).

\begin{figure}[h]
\includegraphics[width=11cm,angle=0]{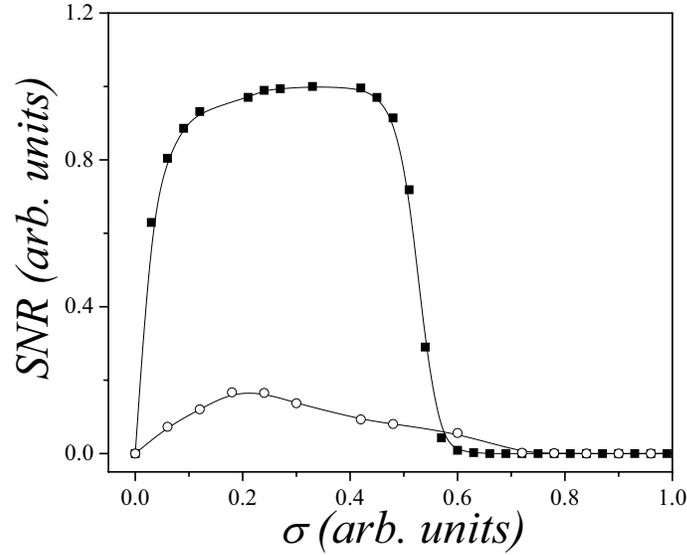}
\caption{SNR obtained when simulating the full set of Eq.
(\ref{x},\ref{y}). Here $\sigma$ corresponds to the noise intensity
defined through the distribution width, as indicated in the text.
Squares and circles corresponds for $\mu=3/2$ and $\mu=5/2$
respectively. The lines are for guiding the eye only.}
\label{fig_snr_exp}
\end{figure}

\begin{figure}[h]
\includegraphics[width=11cm,angle=0]{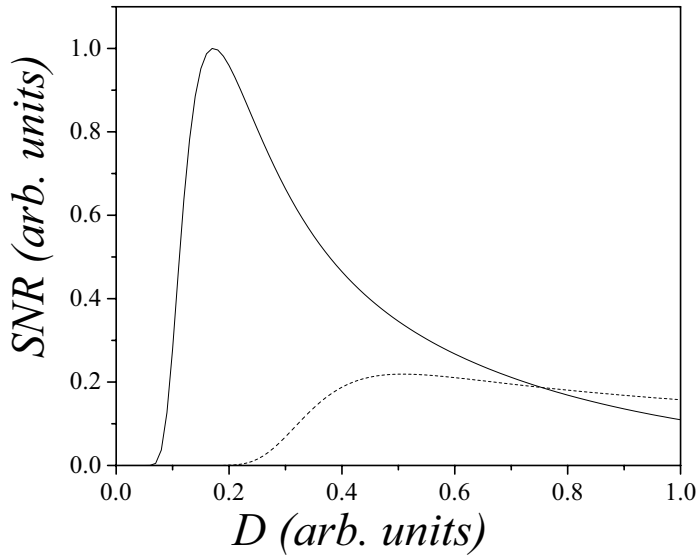}
\caption{SNR obtained using Eq. (\ref{snrtheory}), as derived from
the two state theory. Continue and dashed line correspond to
$\mu=3/2$ and $\mu=5/2$ respectively.} \label{fig_snr_teo}
\end{figure}

\begin{figure}[h]
\includegraphics[width=11cm,angle=0]{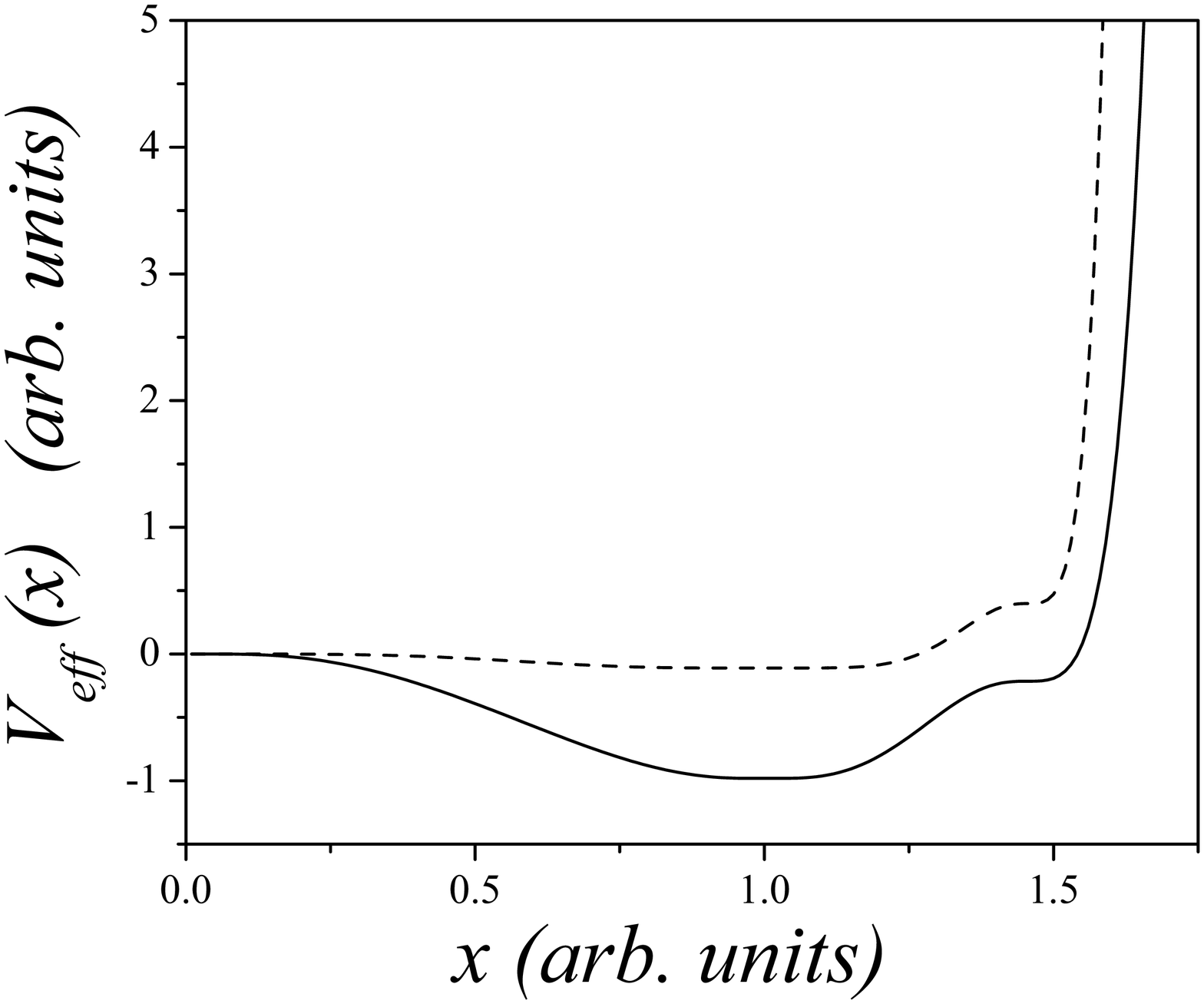}
\caption{Effective potential, Eq (\ref{pot_efectivo}), showing the
different behavior of the wells for different values of the exponent
$\mu$: continuous line $\mu = 5/2$, dotted line $\mu = 3/2$. Due to
the symmetry, we only show positive values of $x$.}
\label{fig_pot_efectivo}
\end{figure}

From the comparison of both figures it is apparent that the results
obtained using the effective Markovian theory are in very good
(qualitative) agreement with those from simulations. This is in
accord with previous results obtained for different systems
\cite{noG,ucna2}. We can conclude that such kind of UCNA-like
approximation offers an adequate framework to obtain effective
Markovian approximations in a very wider class of systems than the
one to which was originally applied \cite{ucna1}.

The above indicated results are in complete agreement with those
of \cite{makra1}. That is: the maximum of the SNR is larger for
white noise, and it moves towards large noise intensities for
increasing $\kappa$. In order to gain some physical insight about
this behavior it is worth to remark that the function defined by
Eq. (\ref{pot_efectivo}) is directly related to an effective
potential within the approximation we used
\begin{eqnarray}
V_{eff}(x) \approx D\, \ln \Psi(x) = D\, \left\{ 2 \int^{x}
\frac{A(\zeta)}{B(\zeta)} d\zeta \right\}. \label{pot_efectivo-2}
\end{eqnarray}
The behavior of such a potential reveals what are the consequences
of changing the exponent $\mu$: when $\mu=5/2$ (i.e. the PSD is
$1/f$), the effective potential shows a well defined well; but as
$\mu$ decreases the well is less defined as shown in Fig. 6. Such a
behavior of the effective potential explains why the SNR increases
when $\mu$ decreases. The general theory shows that the SNR increase
is proportional to the Kramers rate $r_K$, that is given by
\begin{eqnarray}
r_K & = & \frac{1}{\sqrt{2} \pi} \exp{\left\{- \frac{\Delta
V_{eff}}{D} \right\} },
\end{eqnarray}
where $\Delta V_{eff}$ is the high of the barrier in the effective
potential separating the attractors. Hence, the reduction of the SNR
with increasing $\kappa$ (or decreasing $\mu$) could be directly
related with the marked reduction of the barrier separating the
attractors in the effective potential picture. This also explains
the reason for the shift of the SNR maximum towards larger values of
the noise intensity.

In conclusion, we want to remark that this is a first step towards
an analytical understanding of two very important, connected, and
ubiquitous, aspects in natural processes. Those are: the $1/f^k$
behavior of noises' PSD, and its role in signal detection via the SR
mechanism. The physical picture provided by the indicated effective
Markovian approximation, offers an adequate framework to analyze and
understand the main qualitative trends of such phenomenon. Even
more, we expect to apply the same scheme to other noise induced
phenomena when subject to $1/f^k$ noises. This will be the subject
of further work. \\

{\bf Acknowledgedments:} MAF would like to thank the support of
CONICET, Argentina, to the Santa Fe Institute for its support and
hospitality, and to M. Ballard for fruitful discussions related to
the one-dimension FPE description. HSW wants to thank the European
Commission for the award of a \textit{Marie Curie Chair}.


\begin{thebibliography}{99}

\bibitem{gammaitoni} L. Gammaitoni, P. H\"{a}nggi, P. Jung and F.
Marchesoni,  Rev. Mod. Phys. \textbf{70}, 223 (1998).

\bibitem{wellens}  T. Wellens, V. Shatokhin and A. Buchleitner, Rep.
Prog. Phys. \textbf{67}, 45 (2004).

\bibitem{noG} M.A. Fuentes, R. Toral and H.S. Wio, Physica A {\bf
295}, 114 (2001); F.J. Castro, M.N. Kuperman, M.A. Fuentes and H.S.
Wio, Phys. Rev. E {\bf 64}, 051105 (2001); M. A. Fuentes, H. S. Wio
and R. Toral, Physica A {\bf 303}, 91 (2002); M.A. Fuentes, C.
Tessone, H.S. Wio and R. Toral, Fluct. and Noise Letters {\bf 3},
365 (2003).

\bibitem{makra1} P. Makra, Z. Gingl and T. F\"{u}lei, Phys. Lett. A
\textbf{317}, 228 (2003).

\bibitem{nozaki4} I. Hidaka, D. Nozaki and Y. Yamamoto, Phys. Rev.
Lett. \textbf{85}, 3740 (2000).

\bibitem{soma1} R. Soma, D. Nozaki, S. Kwak and Y. Yamamoto, Phys.
Rev. Lett. \textbf{91}, 078101 (2003).

\bibitem{kish1} P. Ruszczynski and L. Kish, Phys. Lett. A
\textbf{276}, 187 (2000).

\bibitem{nozaki1} D. Nozaki, S. Kwak and Y. Yamamoto, Phys. Lett. A
\textbf{243}, 281 (1998).

\bibitem{nozaki2} D. Nozaki, D.J. Mar, P. Grieg and J.J. Collins,
Phys. Rev. Lett. \textbf{82}, 2402 (1999).

\bibitem{nozaki3} D. Nozaki, J.J. Collins and Y. Yamamoto, Phys.
Rev. E  \textbf{60}, 4637 (1999).

\bibitem{kaulakys00} B. Kaulakis and J. Ruseckas, Phys. Rev. E
\textbf{70}, 020101 (2004).

\bibitem{kaulakys01} B. Kaulakis, V. Gontis and M. Alaburda, Phys.
Rev. E \textbf{71}, 051105 (2005).

\bibitem{kaulakys02} B. Kaulakis, J. Ruseckas, V. Gontis and M.
Alaburda, cond-mat/0509626 (2005)

\bibitem{ucna1} P. Jung and P. H\"{a}nggi, Phys. Rev. A \textbf{35},
4464 (1987); L. H'walisz, P. Jung, P. H\"{a}nggi, P. Talkner and L.
Schimanski-Geier, Z. Physik B \textbf{77}, 471 (1989).

\bibitem{ucna2} H.S. Wio, P. Colet, L. Pesquera, M.A. Rodriguez and
M. San Miguel, Phys. Rev. A {\bf 40}, 7312 (1989); P. H\"{a}nggi,
Chem. Phys. \textbf{180}, 157 (1994), F. Castro, A. S\'anchez and
H.S. Wio, Phys. Rev. Lett. \textbf{75}, 1691 (1995); S. Mangioni, R.
Deza, H.S. Wio and R. Toral, Phys. Rev. Lett. \textbf{79}, 2389
(1997).

\bibitem{gardi} C.W. Gardiner,{\it Handbook of Stochastic Methods},
2nd Ed. (Springer-Verlag, Berlin, 1985).

\end{thebibliography}
\end{document}